\documentclass[10pt]{emulateapj}
\begin{document}
\title{Probing the structure of  the outflow in the tidal disruption flare Sw J1644+57 with long-term radio emission  }
\author{Di Cao\altaffilmark{1,2} , Xiang-Yu Wang\altaffilmark{1,2} }
\altaffiltext{1}{Department of Astronomy, Nanjing University,
Nanjing, 210093, China} \altaffiltext{2}{Key laboratory of Modern
Astronomy and Astrophysics (Nanjing University), Ministry of
Education, Nanjing 210093, China}

\begin{abstract}
The recently discovered high-energy transient Sw J1644+57 is thought
to arise from the tidal disruption of a passing star by a dormant
massive black hole. The long-term, bright radio emission of Sw
J1644+57 is believed to result from the synchrotron emission of the
blast wave  produced by an outflow expanding into the surrounding
medium. Using the detailed multi-epoch radio spectral data, we are
able to determine the total number of radiating electrons in the
outflow at different times, and further the evolution of the cross
section of the outflow with time. We find that the outflow gradually
transits from a conical jet to a cylindrical one at later times. The
transition may be due to  collimation of the outflow by the pressure
of the shocked jet cocoon  that forms while the outflow is
propagating in the ambient medium. Since cylindrical jets usually
exist in AGNs and extragalactic jets, this may provide independent
evidence that Sw J1644+57 signals the onset of an AGN.
\end{abstract}


\section{Introduction}
The high-energy transient Swift J164449.3+573451 (hereafter Sw
J1644+57) was detected by Swift Burst Alert Telescope (BAT)  on
March 28, 2011 (Burrows et al. 2011). It was considered as a
gamma-ray burst (GRB) in the beginning. However, Sw J1644+57 has
later shown many different characters from  GRBs and from active
galactic nuclei (AGNs) as well. It has a lifetime much longer than
GRBs and is much more luminous than AGNs. Further observations show
that the position of this event is coincident with the nucleus of a
galaxy at redshift $z \simeq 0.3534$ (Levan et al. 2011; Bloom et
al. 2011). Its long-term X-ray luminosity varies as $L_X \propto
t^{-5/3}$, consistent with the expected fallback rate of the tidally
disrupted material (Rees 1998). Thus this event was later considered
as a star being tidally disrupted by a $ \sim 10^6 M_{\sun}$
supermassive black hole (SMBH) in the center of the galaxy. Since
the $\gamma$-ray and X-ray luminosity of this event is 2-3 orders of
magnitude larger than the Eddington luminosity of a $ \sim 10^6
M_{\sun}$ SMBH, the emission is likely to be produced by a
collimated outflow. It is worth noting that Sw J1644+57 has a bright
long term radio synchrotron emission (Zauderer et al. 2011), which
provide a unique chance for us to study how the physical parameters
of the outflow evolve as it interacts with the ambient medium.

Metzger et al. (2012) modeled the radio-microwave emission during
5-23 days after the trigger with the blast wave emission, in analogy
to GRB afterglows (see e.g. Giannios \& Metzger 2011). They find
that a narrow jet of an opening angle of $\theta_j \lesssim
1/\gamma_j $ expanding into a medium with a density profile $n
\propto R^{-2}$ could fit the radio data. Berger et al. (2012)
reconsidered this model based on much longer radio data during 5-216
days. They noticed that the energy injection model used in Metzger
et al.(2011) is no longer suitable after 23 days. Instead they made
an assumption that the energy distribution in the ejecta is
described by $E(> \Gamma) \propto {\Gamma}^{-2.5}$ (where $\Gamma$
is the bulk Lorentz factor of the ejecta), so that substantial
energy is added into the blast wave at later times. The radial
density profile after 23 days is inferred to be $n \propto R^{-1.5}$
with a plateau at $r \approx {\rm 0.4-0.6pc}$. Both works made a
pre-assumption that the jet has a conical structure with a constant
opening angle $\theta_j \lesssim 1/\gamma_j $. However the jet
structure of this transient has not been  resolved yet. The cross
sections of extragalactic AGN radio jets do not increase at large
radii and cylindrical jet structure (or blob) has been usually
assumed (Bridle \& Perley 1984). Since Sw J1644+57 involves a jet
from a SMBH similar to an AGN, it is necessary to consider the
possibility of a cylindrical jet\footnote{The possibility of
cylindrical jets has also been discussed for GRBs (Cheng et al.
2001; Wang et al. 2005).}.   In this paper we first present an
approach to calculate the parameters of the jet blast wave at any
instantaneous time directly by using the radiation spectrum at that
time (\S 2). This approach applies to both conical and cylindrical
jets. Then, using these parameters, we calculate the jet cross
section of Sw J1644+57 at different times and study how it evolves
as the jet propagates outward (\S 3). We find that the outflow
gradually transits from a conical jet to a cylindrical one at later
times. The possible mechanism for the collimation is further
discussed (\S 3.2). Finally, we give our conclusions (\S 4).

\section{Modeling the radio emission}
In our model, the radio afterglow of Sw J1644+57 is produced by
electrons accelerated by the forward shock that forms as the jet
interacts with the ambient medium. The radiation can be simply
described by a cloud of relativistic electrons radiating synchrotron
emission in the magnetic fields, so it is completely determined by
the following parameters: the total number of  radiating electrons
$N_e$, the bulk motion Lorentz factor $\Gamma$, the magnetic field
$B$, the thermal Lorentz factor of electrons $\gamma_e$ and
power-law distribution index $p$. We made the same assumption that
$\theta_j \lesssim 1/\gamma_j $ in the conical jet case, so the
parameter $\theta_j$ is not involved. The bulk Lorentz factor
relates with the jet energy $E_j$ through
\begin{equation}
\Gamma=\left(\frac{E_j}{m_p N_e c^2}\right)^{\frac{1}{2}}.
\end{equation}
Assuming that the magnetic field energy density is a factor
$\varepsilon_B$ of the shock internal energy, the magnetic field
related with the number density of the ambient medium $n$ through
\begin{equation}
B=\Gamma c (32\pi n m_p \varepsilon_B)^{\frac{1}{2}}.
\end{equation}
The minimum energy of the accelerated power-law electrons is given
by
\begin{equation}
\gamma_m=\varepsilon_e\frac{m_p}{m_e}\frac{(p-2)}{(p-1)}\Gamma,
\end{equation}
where $\varepsilon_e$ is the equipartition factors of electron
energy density.

The radio synchrotron emission produced by the power-law electrons
can be described by three characteristic quantities, i.e. the
typical frequency $\nu_m$, the self-absorption frequency $\nu_a$ and
the peak flux $F_{\rm \nu_m}$ (Sari et al. 1998). The two break
frequencies are, respectively, given by
\begin{equation}
\nu_m=\frac{\Gamma\gamma_m^2 e B}{2\pi m_e c}
\end{equation}
and
\begin{equation}
\nu_a=\nu_m\left(C_0(p-1)\frac{e n
R}{B\gamma_m^5}\right)^{\frac{3}{5}}.
\end{equation}
where $C_0 \approx 10.4\frac{p+2}{p+2/3}$ (Wu et al. 2005). The
emission radius $R$ is given by
\begin{equation}
R=\frac{2\Gamma^2 c t}{1+z},
\end{equation}
where $t$ is the observer time.

In a relativistic cylindrical jet, the radiation emitted by
electrons is distributed over a solid angle of $2\pi/\Gamma^{2}$.
The peak flux can be obtained by calculating the total emission of
radiating electrons $N_e P_{\rm \nu,max}$, where $P_{\rm \nu,max}$
is the peak spectral power (Sari et al. 1998), so the observed peak
flux density is
\begin{equation}
F_{\rm \nu_m}=\frac{N_e P_{\rm \nu,max} (1+z)}{2\pi D_L^2/\Gamma^2}.
\end{equation}
On the other hand, for a conical jet with $\theta_j \lesssim
1/\Gamma$, the observed peak flux density can be written as:
\begin{equation}
F_{\rm \nu_m}=\frac{N_{e,iso} P_{\rm \nu,max} (1+z)}{4\pi D_L^2}
\frac{\Gamma^2 {\theta_j}^2}{2},
\end{equation}
where $N_{e,iso}$ is the equivalent isotropic electron number, which
relates with the real number of electrons $N_e$ by
\begin{equation}
N_e \simeq N_{e,iso} \frac{ {\theta_j}^2}{4}.
\end{equation}
Substituting  $N_{e,iso}$ in equation (9) into equation (8), we find
that the expressions for the peak flux are the same  for both
cylindrical and conical jet models.   Thus equations (4)-(7) apply
to both cylindrical and conical jets.

Equations (4), (5) and (7) can be rewritten as explicit expressions
of five independent parameters $\varepsilon_B$, $\varepsilon_e$,
$N_e$, $n$, and $E_j$, i.e.
\begin{equation}
F_{\rm \nu_m}=C_F\varepsilon_B^{\frac{1}{2}} n^{\frac{1}{2}} E_j^2
N_e^{-1},
\end{equation}
\begin{equation}
\nu_a=C_a\varepsilon_e^{-1}\varepsilon_B^{\frac{1}{5}}
n^{\frac{4}{5}} E_j^{\frac{4}{5}} N_e^{-\frac{4}{5}},
\end{equation}
\begin{equation}
\nu_m=C_m\varepsilon_e^2\varepsilon_B^{\frac{1}{2}} n^{\frac{1}{2}}
E_j^2 N_e^{-2},
\end{equation}
where $C_F$, $C_a$, $C_m$ are three constant coefficients. After
setting typical values for  $\varepsilon_B$ and $\varepsilon_e$ (for
instance, $\varepsilon_B = 0.1$ and $\varepsilon_e = 0.1$;
Panaitescu \& Kumar 2000), we can solve  the other three quantities
$N_e$, $n$, $E_j$, using the radio spectral data at different times.

Once we know the evolution of $N_e$, $n$, $E_j$ with the observer
time $t$, we can obtain $\Gamma(t)$, $R(t)$,  and $n(R)$ with the
assistance of equations (1) and (6).

\section{Parameter evolution of Sw J1644+57}
Now we confront the above model with the observational data of Sw
J1644+57. We adopt the observational radio data of Sw J1644+57 from
Berger et al. (2012).  Fig. 1 shows our fit of the radio spectral
data with the synchrotron broken power-law. We find that the
favorable value for the power-law index is $p \simeq 2.4$. The
self-absorption frequency $\nu_a$ is always below the typical
frequency $\nu_m$ during the observation period.  Table 1 summarizes
the characteristic quantities of the broken power-law spectra at
different times for Sw J1644+57. {With these data, we can get the
characteristic parameters, such as $\Gamma$, $n$ and $R$,  at
different times without assuming the blast wave dynamics beforehand,
i.e. we do not need to assume the ambient density profile or energy
injection mode.}

Because the spectral data during 36-68 days are incomplete,  we
could only get the spectral quantities  $\nu_a$ and $F_{\rm \nu_a}$.
With these quantities the jet energy $E_j$ can be obtained
precisely, while we can only get upper limits for the ambient
density $n$ and the total number of radiating electrons $N_e$.
However, as will be mentioned below, the Lorentz factor of the
ejecta may be constant during 23-216 days. If we take the assumption
that the Lorentz factor during 36-68 days is a constant, e.g.
$\Gamma\simeq3$, the values of $n$ and $N_e$ can be obtained. We
show the evolution of the parameter $E_j(t)$, $n(R)$, $R(t)$ and
$\Gamma(t)$ in figure 2.

\subsection{Evolution of the jet energy and the density profile  of the ambient medium}
The jet energy of Sw J1644+57 evolves as $E_j \propto t^{0.6}$
during 5-23 days (Fig. 2(a)),  $E_j \propto t^{1}$ at later times,
and after 126 days   $E_j \propto t^{0.6}$ again. This indicates
that Sw J1644+57 has a continuous energy injection lasting much
longer than previous thought. The Swift XRT data of Sw J1644+57 obey
the power-law decay $L_x \sim t^{-5/3}$. Such a slow energy
injection is insufficient  to account for the required energy
increase, so an extra or new energy supply mechanism should be
needed. In figure 2(b) we can see that the  ambient density profile
varies as $n \propto R^{-(1.3-1.5)}$, which is consistent with the
prediction by the Bondi accretion as suggested by Berger et al.
(2012). The radius of the jet displayed in figure 2(c) evolves as $R
\propto t^{1.1}$  during 23-216 days and $R \propto t^{0.7}$ during
5-23 days. Since $R \propto \Gamma^{2}t$, this relation indicates
that the Lorentz factor of the ejecta varies as $\Gamma \propto
t^{-0.15}$ during 5-23 days and remains almost constant ($\Gamma
\propto t^{0.05}$) during 23-216 days. This behavior is clearly seen
in figure 2(d). {This is the reason why we assume $\Gamma\simeq3$
during 36-68 days in the calculation.}

\subsection{Jet cross section evolution }
The above equations apply to both conical and  cylindrical jets.
However, the structure could be known if we know how the cross
section of the jet evolves as the radius $R$ increases. The size of
the cross section can be derived from the evolution of $N_e$, $n$
and $R$. Since the total number of swept-up electrons by the jet
relates with the cross section radius through
\begin{equation}
N_e (R) = \int\pi  r_j^2(R) n(R) dR,
\end{equation}
we can derive the cross section radius by
\begin{equation}
r_j (R) = \sqrt{\frac{1}{\pi n(R)}\frac{dN_e}{dR}}.
\end{equation}
To obtain  $\frac{dN_e}{dR}$, we first find an empirical function to
fit the relation between $N_e$ and $R$, and then calculate the
differential coefficient. The result is shown in figure 3. We find
that when  $R$ is smaller than  a few $10^{18}{\rm cm}$, the jet
cross section  evolved as a conical jet with $r_j \propto R$, which
is consistent with the result obtained by Metzger et al. (2012).
However at larger radii ($R\ga4\times10^{18}{\rm cm}$),  $r_j$
increase slowly and it tends to become a constant (i.e. $r_j\sim
{\rm constant}$), which indicates that the jet may transit to  a
cylindrical structure.

{The transition may be due to the  collimation of the jet by the
surrounding cocoon  which forms while the jet is propagating in the
ambient medium (Bromberg et al. 2011). Such a mechanism has been
discussed in long-lived AGN jets (Begelman \& Cioffi 1989) , as well
as in microquasar jets and GRB jets (Bromberg et al. 2011). In this
mechanism, matter in the jet head is heated and flows sideways due
to its higher pressure than the ambient medium, which  leads to the
formation of a pressured cocoon around the jet. If the cocoon
pressure is sufficiently high, it collimates the jet and reduces its
opening angle. Thus a jet can be conical initially and transit to a
cylindrical jet at later times. Bromberg et al. (2011) find the
condition under which the jet will be collimated:
\begin{equation}
\widetilde{L} \simeq \frac{L_j}{\Sigma_j \rho_\alpha c^3}\lesssim
\theta_0^{-4/3},
\end{equation}
where $\widetilde{L}$ is  a dimensionless parameter that defines the
ratio between the energy density of the jet and the rest-mass energy
density of the surrounding medium at the location of the head,
 $L_j$ is the jet luminosity, $\Sigma_j$ is
the jet cross section, $\rho_\alpha$ is the ambient medium density,
and $\theta_0$ is the initial opening angle of jet. This condition
gives a critic transition radius at
\begin{equation}
R_c \gtrsim \sqrt{\frac{L_j}{\pi \rho_\alpha \theta_0^{2/3} c^3}},
\end{equation}
above which the jet becomes collimated and transits to a cylindrical
jet.

For the case of Sw J1644+57, we can obtain the critic transition
radius by taking $L_j=dE_j / dt$, $\rho_\alpha = n m_p$ and assuming
a typical initial opening angle of $\theta_0 =10^{\circ}$. Figure 4
shows the critic radius $R_c$ as a function of time (the dotted
line). It clearly shows the critic radius approaches the jet radius
at late times, which means that jet tends to be collimated at later
times. The transition radius is $\sim 5-8 \times 10^{18} {\rm cm}$,
which roughly (within a factor of a few) agrees with the above
result obtained by modeling of the radio spectral data (Fig. 3). The
fact that the jet in a tidal disruption event transits from conical
to cylindrical structure in a way similar to long-lived AGN jets
provides independent support that Sw J1644+57 may result from the
onset of an AGN.}

\section{Conclusions}
The structure of the jets in tidal disrupted flares such as Sw
J1644+57, whether it is conical as usually assumed in GRB jets or
cylindrical as seen in some extragalactic radio jets, is largely
unknown. We propose that the long-term radio data can be used to
probe the jet structure. By fitting the observed radio data of Sw
J1644+57, we find that the jet structure of Sw J1644+57 undergos a
transition from a conical jet  to a cylindrical one at  tens of days
after the initial flare. It is natural to expect that the jets are
initially conical and later become cylindrical due to collimation by
the surrounding cocoon. Observations of extragalactic radio AGN jets
show that the jets are indeed cylindrical on large scales. The
similar processes occurring in AGNs and Sw J1644+57  may possibly
indicate that Sw J1644+57 also arise from  accretion process in a
supermassive black hole, likely feeded by the tidally captured
stellar material.

We thank Rongfeng Shen for useful discussions and the referee for
the constructive report. This work is supported by the 973 program
under grant 2009CB824800, the NSFC under grants 10973008, 11273016
and 11033002,  the Excellent Youth Foundation of Jiangsu Province
and the Fok Ying Tung Education Foundation.

\begin{table*}
\caption{The radio spectrum fit of Sw J1644+57. Data during 5-23
days are taken from Metzger et al.(2012). The rest are obtained from
the data of Berger et al. (2012) by using the broken power-law
spectral fit.}
\begin{tabular}{lllllllll}
\hline \hline
$t($\rm day$)$ & $\nu_a($\rm GHz$)$ & $\nu_m($\rm GHz$)$ &  $F_{\rm \nu_m}($\rm mJy$)$ & $F_{\rm \nu_a}($\rm mJy$)$\\
\hline
5     & 40   & 386  & 28.6 & -     \\
23    & 11.6 & 50.9 & 15.3 & -     \\
36    & 8.12 & -    & -    & 10.56 \\
51    & 9.34 & -    & -    & 12.93 \\
68    & 8.87 & -    & -    & 17.08 \\
95    & 8.15 & 23.8 & 31.3 & -     \\
126.5 & 7.42 & 23.6 & 30.3 & -     \\
161   & 6.04 & 18.2 & 28.2 & -     \\
197   & 5.38 & 15.3 & 26.5 & -     \\
216   & 5.89 & 17.9 & 26.2 & -     \\
\hline
\end{tabular}
\end{table*}

\begin{figure*}
\plotone{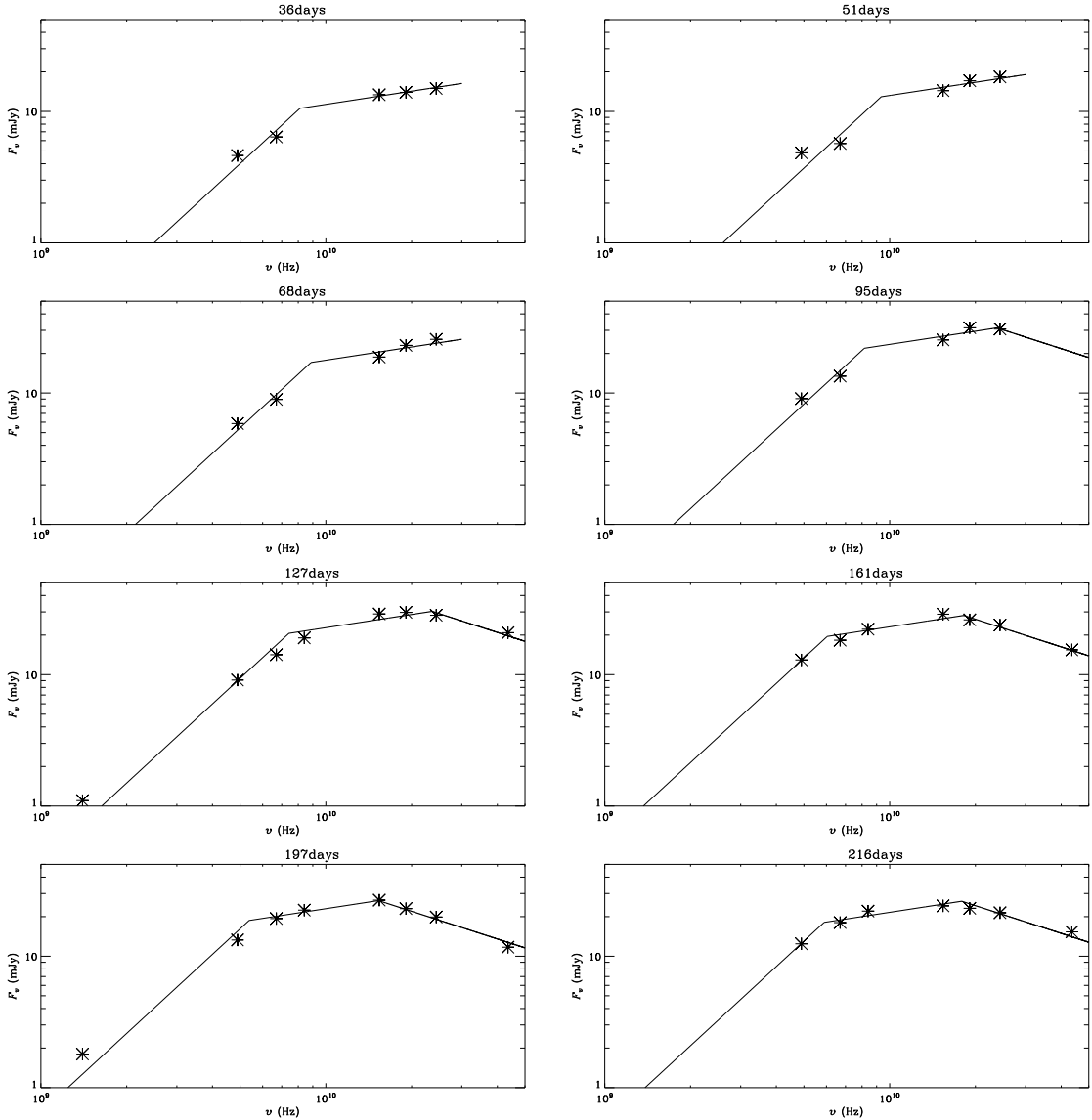} \caption{ The best fit of the radio data  of Sw
J1644+57 using the synchrotron broken power-law spectra. Observation
data are taken from Berger et al. (2012)}
\end{figure*}
\begin{figure*}
\plotone{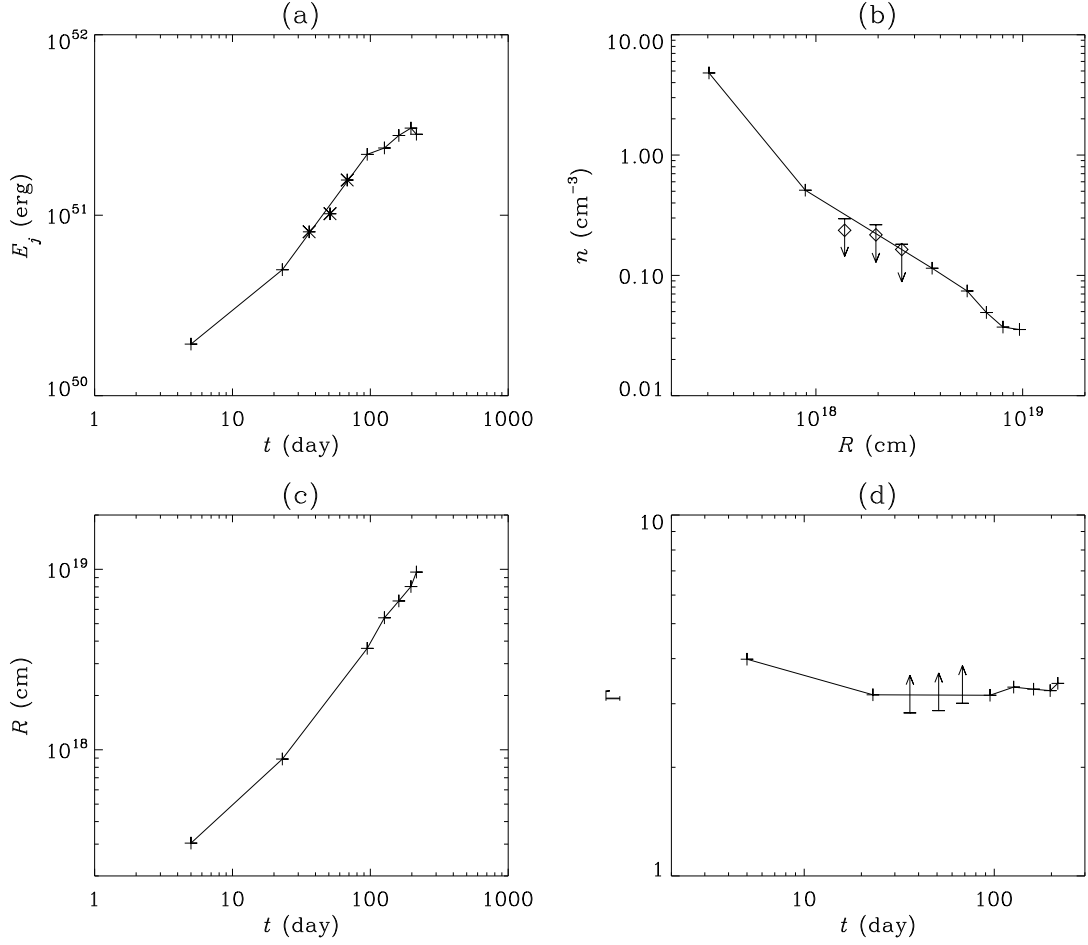} \caption{ Time-evolution of the parameters $E_j$,
$n$, $R$ and $\Gamma$,  corresponding to panels (a)-(d). The
equipartition factors of electron energy density and magnetic
density are $\varepsilon_e = 0.1$ and $\varepsilon_B = 0.1$
respectively. The power-law index is $p = 2.4$. Due to the
incompleteness of radio data during 36-68 days, we plot the
parameters obtained during this period with different symbols. The
star symbol represents  data calculated by using $\nu_a$ and $F_{\rm
\nu_a}$ only. The diamond symbol represents  the  values of the
number density $n$ derived by using the assumption  $\Gamma = 3.16$.
}
\end{figure*}

\begin{figure*}
\plotone{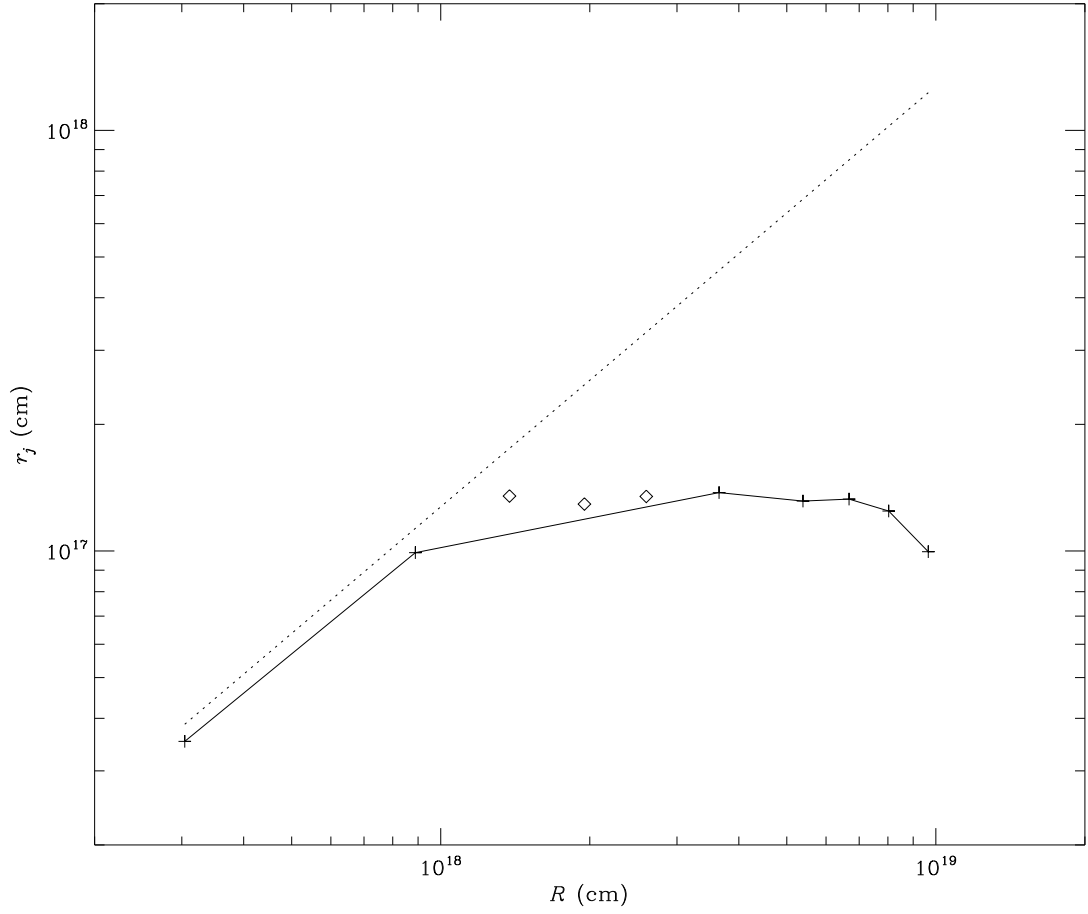} \caption{ Evolution of jet cross section radius
$r_j$. The dotted line exhibits the  expect evolution of $r_j$ for
the conical jet  model. The solid line shows the evolution of $r_j$
derived from  the radio data. }
\end{figure*}

\begin{figure*}
\plotone{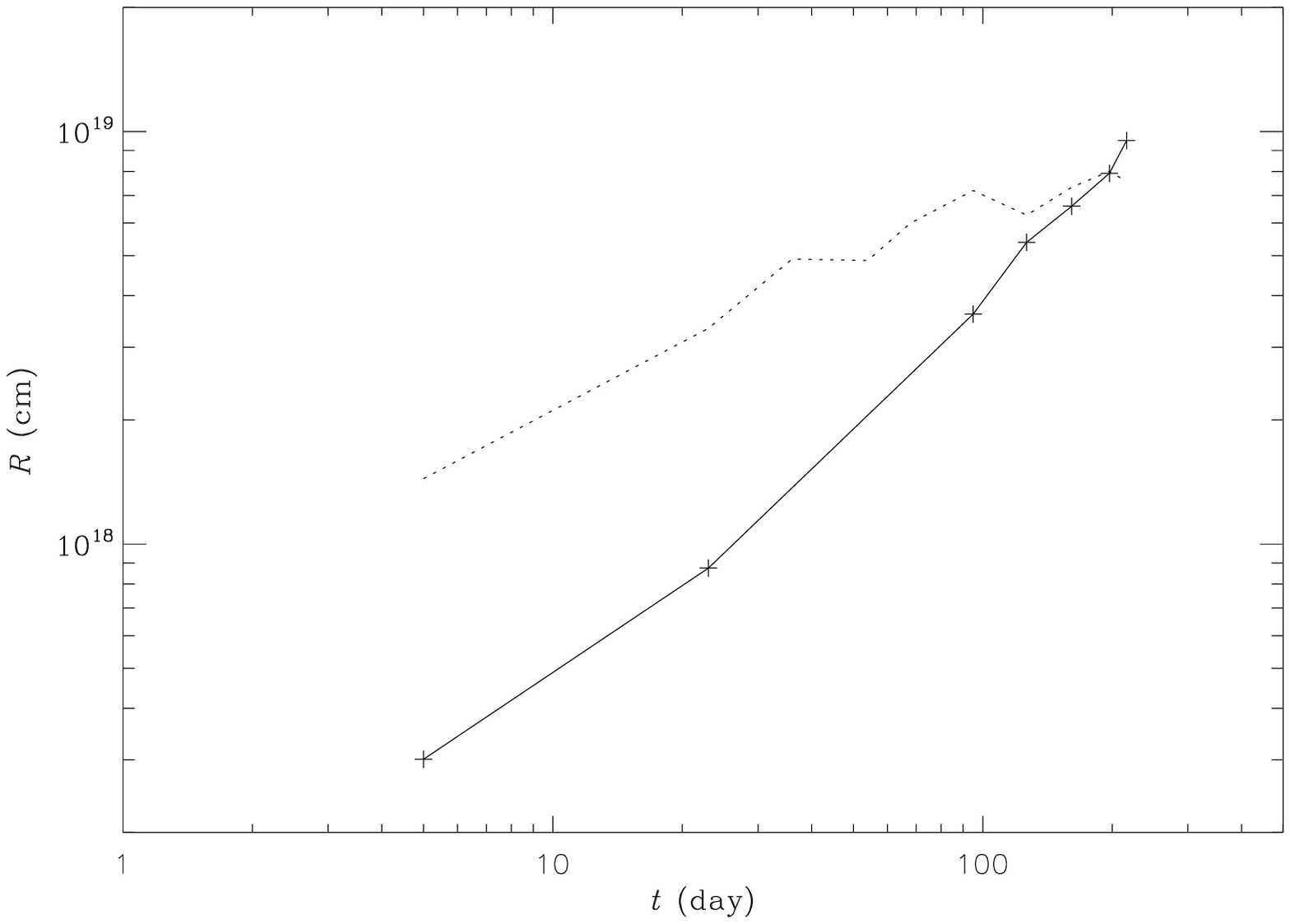} \caption{The dotted line denotes the critic
radius  of jet  collimation for the parameters of Sw J1644+57,
obtained with the condition found by Bromberg et al (2011). The
solid line shows the evolution of the jet radius $R$ inferred from
the observed radio data.}
\end{figure*}


\begin{thebibliography}{99}

\bibitem{}
Berger, E. et al. 2012, ApJ, 748, 36

\bibitem{}
Beskin, V. S. 1997,  Phys.-Uspekhi, 40 (7), 659

\bibitem{}
Bloom, J. S. et al. 2011, Science, 333, 203
\bibitem{}
Bridle, A. H., \& Perley, R. A. 1984, ARA\&A, 22, 319

\bibitem{}
Bromberg, O. et al. 2011, Apj, 740, 100

\bibitem{}
Burrows, D. N. et al. 2011, Nature, 476, 421

\bibitem{}
Cheng, K. S., Huang, Y. F. \& Lu, T. 2001, MNRAS, 325, 599


\bibitem{}
Giannios, D. \& Metzger, B. D. 2011, MNRAS, 416, 2102

\bibitem{}
Levan, A. J. et al. 2011, Science, 333, 199

\bibitem{}
Metzger, B. D. et al. 2012, MNRAS, 420, 3528

\bibitem{}
Panaitescu, A. \& Kumar, P. 2000, ApJ, 543, 66

\bibitem{}
Rees, Martin J. 1988, Nature, 333, 523

\bibitem{}
Sari, R., Piran, T. \& Narayan, R. 1998, ApJ, 497L, 17

\bibitem{}
Wang, X. Y., Cheng K. S. \& Tam P. H. 2005, ApJ, 621, 894

\bibitem{}
Wu, X. F. et al. 2005, ApJ, 619, 968

\bibitem{}
Zauderer, B. A. et al. 2011, Nature, 476, 425

\end{thebibliography}
\end{document}